\documentclass[12pt]{article}
\usepackage{geometry}                
\usepackage{graphicx}
\usepackage{amssymb}
\usepackage{epstopdf}
\newlength{\diagwidth}
\newcommand{\den}[0]{{\rm den}}
\newcommand{\text}[1]{{\rm #1}}

\title{
The Social Will-Testing Game \\and its Solution}
\author{Leonid Gurvits and J. Stephen Judd}

\begin{document}
\maketitle
\abstract{We examine a two-person game we call Will-Testing in which the strategy space for both players is a real number. It has no equilibrium.
When an infinitely large set of players plays this in all possible pairings, there is an equilibrium for the distribution of strategies which requires all players to use different strategies. We conjecture this solution could underlie some phenomena observed in animals.
}

\section{The Will-Testing Game}
This game is played by two people in continuous time through a fixed time interval. Either player may capitulate at any time, thereby ending the play. The payoff for the non-capitulator is some constant times the amount of time remaining in the game; the capitulator gets half as much as the non-capitulator.

The tension in the game is that both players want someone to capitulate in a hurry,
but both want the other one to do so, and hence they both stall.
There is no equilibrium to this game. The maximum sum of payoffs is $1 + .5$, so players could judge their earnings by how much of that they obtained; many people would judge $0.75$ to be a satisfactory take-away pay.
In the case where this game is played repeatedly with the same player, there is an opportunity to split the $1.5$ quite evenly in half, but achieving this requires the development of trust.
How humans or chickens or other animals play this game is an area open to behavioral studies.

The tension changes as we alter the fraction that the capitulator gets.
Clearly if his fraction is 1, then it does not matter who capitulates first, and both will do so immediately. If his fraction very small (say 1/100) then there is a huge incentive to stall.
If his fraction is close to 1 then it does not matter very much who goes first, and it's relatively
easy to make the sacrifice to go first. Since the fraction makes a big difference, we
will parameterize the game and call the capitulator's fraction $\rho$ where $0\le \rho \le1$.

In spite of its description as being played in continuous time, we will formalize it as a one-shot game where each player $i$ selects a real number $s_i$ in the fixed interval $[0,T]$. 
Then the payoffs for players 1 and 2 respectively are:

\[\left\{\begin{array}{cccl}
(T-s_2)& \mbox{ and } &\rho\times(T-s_2) &  \mbox{if $s_1>s_2$}\\
\rho\times(T-s_1)& \mbox{ and } &\rho\times(T-s_1) & \mbox{if $s_1=s_2$}\\
\rho\times(T-s_1)& \mbox{ and } &(T-s_1) & \mbox{if $s_1<s_2$}\\
\end{array}\right.\]

The {\em social version} of the game has a large set of players who each choose a strategy once; 
their payoffs are the average of the payoffs using that strategy against all other players. 
This captures the idea of a group of social animals that play the game repeatedly with 
partners picked uniformly at random from the same group.

There is no fixed pure equilibrium for this game when there are a finite number of players.
In the limit of infinite players there is a density function for the equilibria, but it does not
specify which player uses which strategy so all permutations are equivalent. 
Finding this distribution involves an equation that makes the usual assumption that all players get equal payoffs, so the players are quite indifferent to which permutation occurs.

One strong interesting feature of the equilibria is that all players must choose {\em different} strategies.
Birds and mammals and many other animals are generally perceived as being 
individuals who are all slightly different, populating a spectrum of behavior space. 
They are also widely seen as having social structures like dominance hierarchies 
wherein pairwise relations between individuals can be characterized. 
We view the social version of Will-Testing to be an example of a game that produces a similar ordered structure where the existence of differences in individual behaviors is a straightforward requirement of the game's solution.
In the case of this particular game, the social dynamic will be interpretable as a total ordering 
(like a pecking order) simply because the strategy space is real and one dimensional.

\section{Deriving the Equilibria}

We seek a density over the strategy space, $\den(t)\ge 0$ and $ \int_0^T\den(t)dt=1$ .
To find the average over all possible pairwise interactions in the social version of the game, 
we must integrate over that space. Define the pay of strategy $s$ as:
\begin{equation}\label{setup}
\text{pay}(s) = \int_0^s \text{den}(t) (T-t) \, dt   +  \int_s^T \rho\;  \text{den}(t) (T-s) \,   dt
\end{equation}

Now make the usual assumption about the nature of the equilibrium that no strategy
is more profitable than any other (otherwise some player would choose that strategy rather
than the one he did), i.e. let pay be a constant:
\[
\text{pay}(s) = c = \text{pay}(0) =  \int_0^T \rho\;  \text{den}(t) (T-0) \,   dt = 
\rho\; T \int_0^T  \text{den}(t) \,   dt = \rho T 
\]
The pay per time spent is thus just $\rho$ for every player, 
which is what the capitulator gets if he acts immediately.

Differentiating both sides of (\ref{setup}), we get a simplified integral equation
\begin{equation} 0= 
\text{den}(s) (T-s)
-\rho\;  \text{den}(s)  (T-s)
-\rho\; \int_s^T  \text{den}(t) \, dt   
\end{equation}
One more differentiation gives the following DE:
\begin{equation}  
 0 =  (1-\rho) (T-s) \text{den}'(s)+(2 \rho -1) \text{den}(s) 
\end{equation}  
\begin{equation}  
	\frac{\text{den}'(s)}{\text{den}(s) } = \frac{-(2 \rho -1)}{(1-\rho) (T-s)} 
\end{equation}
The solution to that is
\begin{equation}  
 \text{den}(s)= \frac{(T-s)^\gamma}{\alpha},\;\;\;
\gamma=\frac{2\rho-1}{1-\rho}
\end{equation}  
and its normalizing factor is
\[ \alpha= \frac{(1-\rho) T^{\frac{\rho }{1-\rho }}}{\rho } \]
which, for the case $T=1$, is simply $ \alpha= (1-\rho)/\rho $.

The integrability condition spells out as:
\begin{equation} \frac{2\rho-1}{1-\rho}>-1 \end{equation}
which gives the following bounds on the parameter $\rho$: $0<\rho<1$.

\begin{figure}[htb] 
   \centering
   \includegraphics[width=5in]{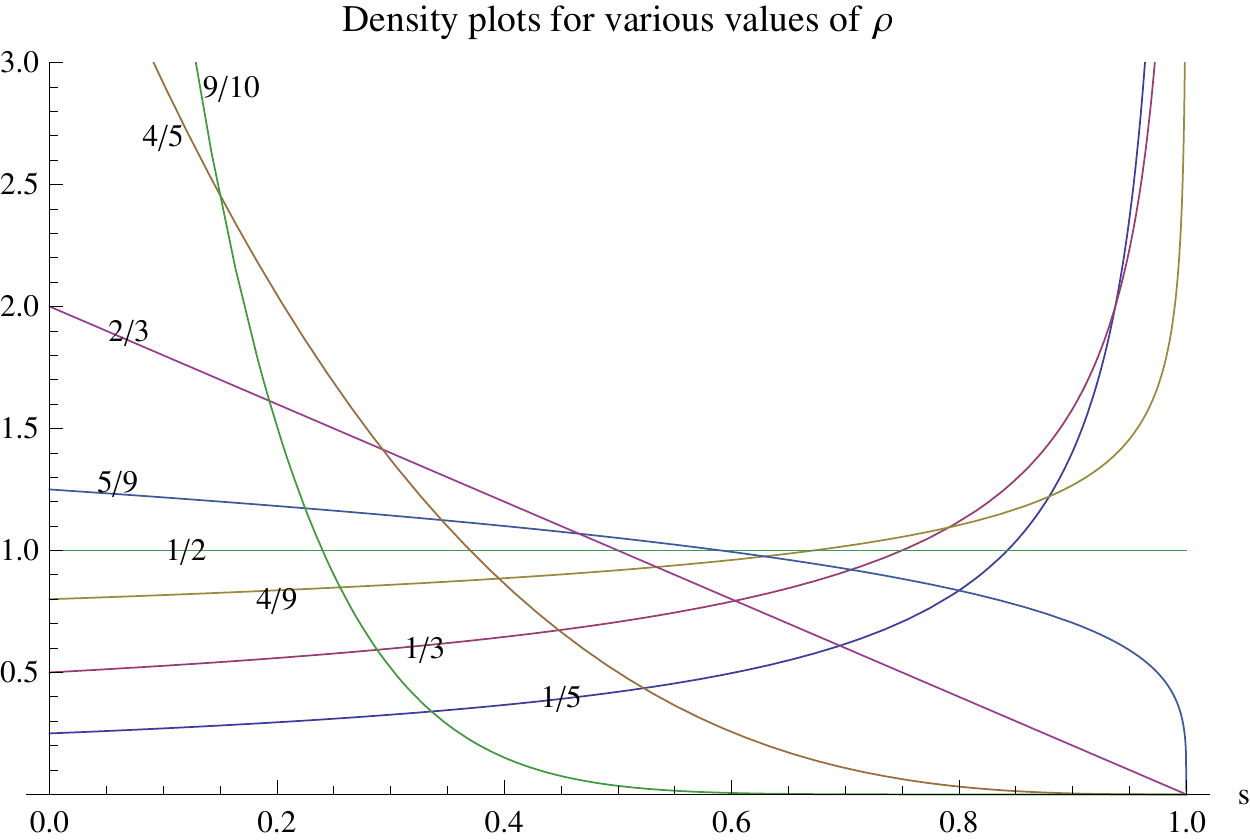} 
   \caption{Equilibria distributions of $s$ for various values of $\rho$ plotted for T=1.
   The case of $\rho=1/2$ is uniform. Any $\rho$ value greater than $1/2$ encourages early capitulation and its density at $s=T$ is zero.
   Any value less than $1/2$ encourages late capitulation and its density at $s=T$ diverges.
      The case of $\rho=2/3$ is the only other straight line. }
   \label{fig:densities}
\end{figure}

Several example distributions are plotted in Figure \ref{fig:densities}.

\section{Discussion}
We have provided a solution to the social version of the Will-Testing game. It fully specifies the equilibria modulo permutations. The trick to solving the equation was to take two consecutive derivatives.

Variation in behavior within an interacting set of animals might have multiple explanations, but we have not seen such variance explained from a game-theoretic standpoint before. In this paper, we have provided a social version of a simple game that could easily be the foundation of a real-world game, and shown that it has a solution of the sort that echoes observed phenomena like pecking orders. Behavioral studies of this game would be informed by comparison with our solution.

The element of time in the Will-Testing game is not essential. Time stands as a proxy for any resource that the players are willing to spend in order to convince the other player to capitulate. Physical struggles that cost energy or other contests that risk bodily health are just as suitable descriptions.

As described, players pick a strategy from $[0,T]$ and use it consistently; revision of this choice is a philosophical event that happens only in the heads of game theorists who analyze it. Another equally valid interpretation is that players change their value of $s$ before every game, and they use the mixed strategy of choosing $s$ randomly according to the distributions found. 

What happens if two players meet each other repeatedly and are allowed to revise their choice of strategy before each game? This is the simple Will-Testing game. By playing a mixed strategy that employs the distribution given above, one player can guarantee that the other one will receive an expected payoff of $\rho$. By playing $s=T$, a player can guarantee that an opponent using the mixed strategy will receive an expected payoff of $\rho/2$. By playing $s=0$, a player can guarantee that his opponent will receive a payoff of $1$. By alternating between $s=0$ and $s>0$, two players can split the max social welfare quite fairly between themselves. By playing $s=T$, a player can guarantee that his opponent will always receive the fraction $\rho$ of what he gets himself. These threats and opportunities lead to interesting dynamics, but no Nash equilibrium exists.

Our solution is for the limit case of infinite population. When the group is finite, then let $S$ be the strategy set that lists all players' strategies. The $\text{pay}_S(s)$ function is a saw function that is discontinuous at every point in $S$, and linearly decreasing in between them. If $S$ is fully observable, then there is no pure equilibrium because many players will discover and switch to a better choice than the one they currently had; that leads to a change in $S$ for everyone else, and constant churning. In the infinite population limit, the segments of this function become infinitesimally narrow and the advantage of being near the low end of them disappears to zero.

If $S$ is not closely observable, then the opportunity to cherry-pick the best alternative strategies is not as potent. Posit that players are allowed to change their strategy at a certain rate. The more players there are, the faster $S$ can change. And as the size of $S$ grows, it takes longer to sample the distribution to gain sufficient information to know where the best choices are. For this reason, we believe that the limit distribution becomes relevant even for modest set sizes. As the set gets bigger, the sampling time goes up while the stability goes down; at some point the players cannot sample fast enough to track the movement, and they have to rely on long-term estimates. From an evolutionary viewpoint then, it might be easier to avoid the detailed computation and simply produce a set of personality traits in the community that approximate the game's limit equilibria.

Note that the nature of the social version depends on uniform sampling of the population. If players could control who they interact with, they could effectively reduce the size of $S$ far enough to avoid the limit, and start playing the simple version instead. For the case of slowly moving animals situated in a large geographic landscape, our definition of the social version may not be relevant. Some network of possible interactions would constrain the choices, and it would be useful to study how the structure of such networks might influence the outcome. The social version defined here is merely a special case in which the interaction network is the complete graph.

We note that we have not been specific as to whether the strategy of both the players is always observable. The capitulator's value is always knowable since it is calculable from the payoff. But it might be that the other player's strategy is not communicated to the capitulator. If so, an asymmetry of knowledge exists. Some variants of Will-Testing might be sensitive to this.

\end{document}